\documentclass[10pt]{revtex4}
\usepackage{amsfonts}
\usepackage{amssymb,epsf}
\usepackage{latexsym}
\begin{document}  

\title{Lovelock gravity from entropic force}
\author{A.  Sheykhi$^{1,2}$, \footnote{
sheykhi@uk.ac.ir} H. Moradpour $^{1}$ and N. Riazi$^{1}$
\footnote{ riazi@susc.ac.ir}}
\address{$^1$ hysics Department and Biruni Observatory, College of
Sciences, Shiraz University, Shiraz 71454, Iran\\
         $^2$ Research Institute for Astronomy and Astrophysics of Maragha (RIAAM), P.O. Box 55134-441, Maragha, Iran}

\begin{abstract}
In this paper, we first generalize the formulation of entropic
gravity to ($n+1$)-dimensional spacetime. Then, we propose an
entropic origin for Gauss-Bonnet gravity and more general Lovelock
gravity in arbitrary dimensions. As a result, we are able to derive
Newton's law of gravitation as well as the corresponding Friedmann
equations in these gravity theories. This procedure naturally leads
to a derivation of the higher dimensional gravitational coupling
constant of Friedmann/Einstein equation which is in complete
agreement with the results obtained by comparing the weak field
limit of Einstein equation with Poisson equation in higher
dimensions. Our study shows that the approach presented here is
powerful enough to derive the gravitational field equations in any
gravity theory. \\
keywords: entropic; gravity; thermodynamics.
\end{abstract}

\maketitle

\section{Introduction}
Nowadays, it is a general belief that there should be some deep
connection between gravity and thermodynamics. Indeed, this
connection has a long history since the discovery of black holes
thermodynamics in 1970's by Bekenstein and Hawking \cite{HB}. The
studies on the profound connection between gravity and
thermodynamics have been continued \cite{B,D} until in 1995 Jacobson
\cite{Jac} disclosed that the Einstein field equation is just an
equation of state for spacetime and in particular it can be derived
from the the first law of thermodynamics together with the relation
between the horizon area and entropy. Inspired by Jacobson's
arguments, an overwhelming flood of papers has appeared which
attempt to show that there is indeed a deeper connection between
gravitational field equations and horizon thermodynamics. It has
been shown that the gravitational field equations in a wide variety
of theories, when evaluated on a horizon, reduce to the first law of
thermodynamics and vice versa. This result, first pointed out in
\cite{Pad}, has now been demonstrated in various theories including
f(R) gravity \cite{Elin}, cosmological setups
\cite{Cai2,Cai3,CaiKim,Wang,Cai33,Shey0}, and in braneworld
scenarios \cite{Shey1,Shey2}. For a recent review on the
thermodynamical aspects of gravity and complete list of references
see \cite{Pad0}. The deep connection between horizon thermodynamics
and gravitational field equations, help to understand why the field
equations should encode information about horizon thermodynamics.
These results prompt people to take a statistical physics point of
view on gravity.

A remarkable new perspective was recently suggested by Verlinde
\cite{Ver} who claimed that the laws of gravitation are no longer
fundamental, but rather emerge naturally from the second law of
thermodynamics as an ``entropic force". Similar discoveries are also
made by Padmanabhan \cite{Padm} who observed that the equipartition
law for horizon degrees of freedom combined with the Smarr formula
leads to the Newton's law of gravity. This may imply that the
entropy links general relativity with the statistical description of
unknown spacetime microscopic structure when a horizon is present.
The investigations on the entropic gravity has attracted a lot of
interest recently \cite{Cai4,Smolin,Li,Tian,
Vancea,Modesto,Sheykhi1,BLi,Sheykhi2,Gu,other,mann}.

On the other hand,  the effect of string theory on classical
gravitational physics is usually investigated by means of a low
energy effective action which describes gravity at the classical
level. This effective action consists of the Einstein-Hilbert action
plus curvature-squared (Gauss-Bonnet) term and also higher order
derivatives curvature terms. Lovelock gravity \cite{Lov,Der} which
is a natural generalization of Einstein gravity in higher
dimensional spacetimes contains higher order derivatives curvature
terms, however there are no terms with more than second order
derivatives of metric in equations of motion just as in Gauss-Bonnet
gravity. Since the Lovelock tensor contains metric derivatives no
higher than second order, the quantization of the linearized
Lovelock theory is ghost-free \cite{Zwi}.

Since the entropic gravity is fundamentally based on the holographic
principle, one expects that entropic gravity can be generalized to
any arbitrary dimension \cite{Ver}. The motivation for studying
higher dimensional gravity originates from string theory, which is a
promising approach to quantum gravity. String theory predicts that
spacetime has more than four dimensions. Another striking motivation
for studying higher dimensional gravity comes from AdS/CFT
correspondence conjecture \cite{Mad}, which associates an
$n$-dimensional conformal field theory with a gravitational theory
in $(n+1)$ dimension. The generalization of this duality is embodied
by the holographic principle \cite{Sus}, which posits that the
entropy content of any region of space is defined by the bounding
area of the region. These considerations have provided us enough
motivation to study the formulation of the entropic gravity in
$(n+1)$-dimensional spacetime. In this paper, we consider the
problem of formulating entropic gravity in all higher dimensions. We
also show that in an string inspired model of gravity the formalism
of entropic force works well and can be employed to derive the
Newton's law of gravity as well as the $(n+1)$-dimensional Friedmann
equation in Gauss-Bonnet theory and more general Lovelock gravity.

This paper is organized as follow. In the next section we generalize
the entropic gravity to arbitrary dimensions and will derive
successfully Newton's law of gravitation as well as Friedmann
equation in $(n+1)$-dimensions. In section III, we derive Newton's
law of gravity and the $(n+1)$-dimensional Friedmann equation in
Gauss-Bonnet theory from the entropic gravity perspective. In
section IV, we generalize our study to the more general Lovelock
gravity. The last section is devoted to conclusions and discussions.

\section{Entropic Gravity in ($n+1$)-dimensions}
According to Verlinde, when a particle is on one side of screen and
the screen carries a temperature, it will experience an entropic
force equal to
\begin{equation}\label{F}
F=-T\frac{\triangle S}{\triangle x}.
\end{equation}
By definition, $F$ is a force resulting from the tendency of a
system to increase its entropy. Note that $\triangle S>0$ and hence
the sign of the force is determined by how one chooses the
definition of $\triangle x$ as it relates to the proposed system.
Here $\triangle x$ is the displacement of the particle from the
holographic screen, while $T$ and $\triangle S$ are the temperature
and the entropy change on the screen, respectively. Suppose we have
a mass distribution $M$ which is distributed uniformly inside an
screen $\Sigma$. We have also a test mass $m$ which is located
outside the screen. The surface $\Sigma$ surrounds the mass
distribution $M$ has a spherically symmetric property, while the
test mass $m$ is assumed to be very close to $\Sigma$ comparing to
its reduced Compton wavelength $\lambda_m=\frac{\hbar}{mc}$. Now,
consider an ($n+1$)-dimensional spacetime with $n$ spacial
dimensions. The mass $M$ induces a holographic screen $\Sigma_n$ at
distance $R$ that has encoded on it gravitational information. The
volume and area of this $n$-sphere are
\begin{equation}
V_n=\Omega_n R^n,  \  \   \   \Sigma_n=n\Omega_{n}R^{n-1},
\end{equation}
where
\begin{equation}
\Omega_{n}=\frac{\pi^{n/2}}{\Gamma(\frac{n}{2}+1)},  \  \  \   \
\Gamma(\frac{n}{2}+1)=\frac{n}{2} (\frac{n}{2}-1)!.
\end{equation}
According to the holographic principle, the screen encodes all
physical information contained within its volume in bits on the
screen. The maximal storage space, or total number of bits, is
proportional to the area $\Sigma_n$. Let us denote the number of
used bits by $N$. It is natural to assume that this number is
proportional to the area $\Sigma_n$, namely
\begin{equation}\label{N}
\Sigma_n=NQ,
\end{equation}
where $Q$ is a constant which should be specified later. Since $N$
denotes the number of bits, thus for one unit change we find
$\triangle N=1$. Therefore, from relation (\ref{N}) one gets
$\triangle \Sigma_n=Q$. Motivated by Bekenstein's area law of black
hole entropy, we assume the entropy of the ($n-1$)-dimensional
holographic screen obeys the area law, namely
\begin{equation}\label{s0}
S=\frac{k_B c^3 \Sigma_n}{4\hbar G_{n+1}},
\end{equation}
where
\begin{equation}\label{G}
G_{n+1}=2 \pi^{1-n/2}\Gamma(\frac{n}{2})\frac{c^3\ell _p^{n-1}
}{\hbar},
\end{equation}
is the ($n+1$)-dimensional gravitational constant \cite{mann}. We
also assume the entropy change
\begin{equation}\label{dS}
\triangle S=\frac{k_B c^3 \triangle \Sigma_n}{4\hbar
G_{n+1}}=\frac{k_B c^3 Q}{4\hbar G_{n+1}}.
\end{equation}
is one fundamental unit of entropy when $\triangle
x=\frac{\hbar}{mc}$, and the entropy gradient points radially from
the outside of the surface to inside. Assuming that the total energy
of the system,
\begin{equation}
\label{Ec} E=M c^2,
\end{equation}
is evenly distributed over the bits. Then according to the
equipartition law of energy \cite{Pad1}, the total energy on the
screen is
\begin{equation}
\label{E}
E=\frac{1}{2}Nk_B T.
\end{equation}
Combining Eqs. (\ref{N}), (\ref{Ec}) and (\ref{E}), we find
\begin{equation}\label{T}
T=\frac{2Mc^2 Q}{\Sigma_nk_B }.
 \end{equation}
Finally, inserting Eqs.  (\ref{dS}) and (\ref{T}) as well as
relation $\triangle x=\frac{\hbar}{mc}$ in Eq. (\ref{F}), after
using relation $\Sigma_n=n\Omega_{n}R^{n-1}$, it is straightforward
to show that the entropic force yields the ($n+1$)-dimensional
Newton's law of gravitation
\begin{equation}\label{F3}
F=-\frac{Mm}{R^{n-1}}\left[\frac{Q ^2c^6}{2n \Omega_n \hbar^2
G_{n+1}}\right],
\end{equation}
This is nothing but the Newton's law of gravitation in arbitrary
dimensions provided we define
\begin{equation}\label{Q}
Q^2\equiv\frac{2\hbar^2}{c^6}n \Omega_n  G^{2}_{n+1}.
\end{equation}
For $n=3$ we have $G_4=G=\ell _p^2c^3 /\hbar$ and the above
expression reduces to $Q^2=8\pi \ell _p^4$ \cite{Sheykhi1}. It is
important to note that the relations $N =\frac{A c^3}{G \hbar}$
and $\triangle S= 2\pi k_B $ postulated by Verlinde \cite{Ver}.
Our assumption here differs a bit from Verlinde postulates. For
example, we have taken $\triangle S=k_B\frac{c^3 Q}{4\hbar G}$
instead of $\triangle S= 2\pi k_B $. Indeed, $\triangle S$ in our
paper is $\frac{1}{\sqrt {8\pi}}\times\triangle S$ Verlinde. So we
do not expect to have exactly $Q=\frac{G\hbar}{c^3}$. Although our
main assumptions $\Sigma_n \propto N$ and $\triangle S \propto k_B
$ are similar but the constants of proportionality are just
assumption for later convenience [16]. Combining Eq. (\ref{Q})
with (\ref{F3}) we reach
\begin{equation}\label{F4}
F=-G_{n+1}\frac{Mm}{R^{n-1}}.
\end{equation}
As the next step, we generalize the study to the cosmological setup.
Assuming a homogeneous and isotropic Friedmann-Robertson-Walker
(FRW) spacetime which is described by the line element
\begin{equation}
ds^2={h}_{\mu \nu}dx^{\mu} dx^{\nu}+R^2d\Omega_{n-1}^2.
\end{equation}
Here $R=a(t)r$, $x^0=t, x^1=r$, and $h_{\mu \nu}$=diag $(-1,
a^2/(1-kr^2))$ is the two dimensional metric, while
$d\Omega^{2}_{n-1}$ is the metric of $(n-1)$-dimensional unit
sphere. The dynamical apparent horizon can be determined using
relation $h^{\mu \nu}\partial_{\mu}R\partial_{\nu}R=0$. It is a
straightforward calculation to show that the radius of the apparent
horizon for the FRW universe becomes
\begin{equation}
\label{radius}
 R=ar=\frac{1}{\sqrt{H^2+k/a^2}}.
\end{equation}
We also assume the matter source in the FRW universe is a perfect
fluid with stress-energy tensor
\begin{equation}\label{T2}
T_{\mu\nu}=(\rho+p)u_{\mu}u_{\nu}+pg_{\mu\nu}.
\end{equation}
Conservation of energy-momentum in $(n+1)$-dimensions leads to the
following continuity equation
\begin{equation}\label{Cont}
\dot{\rho}+n H(\rho+p)=0,
\end{equation}
where $H=\dot{a}/a$ is the Hubble parameter. First of all, we derive
the dynamical equation for Newtonian cosmology. Consider a compact
spatial region $V_n$ with a compact boundary $\Sigma_n$, which is a
sphere with physical radius $R= a(t)r$. If we combine the
gravitational force (\ref{F4}) with the second law of Newton for the
test particle $m$ near the screen $\Sigma_n$, then we obtain
\begin{equation}\label{F6}
F=m\ddot{R}=m\ddot{a}r=-G_{n+1}\frac{Mm}{R^{n-1}}.
\end{equation}
The total physical mass $M$ in the spatial region $V_n$ is defined
as \cite{Cai4}
\begin{equation}\label{M0}
M=\int{dV \left(T_{\mu\nu}u^{\mu}u^{\nu}\right)}=\Omega_n R^{n}
\rho,
\end{equation}
where $\rho=M/V_n$ is the energy density of the matter inside the
the volume $V_n=\Omega_n R^{n}$. Combining Eqs. (\ref{F6}) and
(\ref{M0}) we reach
\begin{equation}\label{F7}
\frac{\ddot{a}}{a}=-G_{n+1}\Omega_n
\rho=-\frac{2G_{n+1}\pi^{n/2}}{n(\frac{n}{2}-1)!}\rho.
\end{equation}
This is the dynamical equation for ($n+1$)-dimensional Newtonian
cosmology. In four dimensional spacetime where $n=3$, we recover the
well-known formula,
\begin{equation}\label{F8}
\frac{\ddot{a}}{a}=-\frac{4 \pi G}{3}\rho.
\end{equation}
In order to derive the ($n+1$)-dimensional Friedmann equations of
the FRW universe, let us notice that the quantity which produces the
acceleration in a dynamical background is the active gravitational
mass $\mathcal M$ \cite{Pad3} rather than the total mass $M$. To
determine the active gravitational mass, we should express $\mathcal
M$ in terms of energy-momentum tensor $T_{\mu\nu}$. The key point
here is to connect the energy momentum $T_{\mu\nu}$ with the
spacetime curvature with the use of the Tolman-Komar's definition of
active gravitational mass. The active gravitational mass in
$(n+1)$-dimension is defined as \cite{Cai4}
\begin{equation}\label{M1}
\mathcal M =\frac{n-1}{n-2}
\int_V{dV\left(T_{\mu\nu}-\frac{1}{n-1}Tg_{\mu\nu}\right)u^{\mu}u^{\nu}}.
\end{equation}
It is a matter of calculation to show that
\begin{equation}\label{M2}
\mathcal M =\frac{\Omega_n R^n}{n-2}[(n-2)\rho+np]=\frac{2\pi^{n/2}
R^n}{n(n-2)(\frac{n}{2}-1)!}[(n-2)\rho+np].
\end{equation}
Now, we can combine Eq. (\ref{M2})  with  (\ref{F6}) provided we
replace $ M$ in Eq. (\ref{F6}) with induced active gravitational
mass $\mathcal M$. This can be done because according to the weak
equivalence principle of general relativity, the active
gravitational mass of a system (here the universe) in general
relativity is equal to its total mass in Newtonian gravity. We find
\begin{equation}\label{addot}
\frac{\ddot{a}}{a} =-\frac{G_{n+1}
}{n-2}\Omega_n[(n-2)\rho+np]=-\frac{2G_{n+1}\pi^{n/2}
}{n(n-2)(\frac{n}{2}-1)!}[(n-2)\rho+np].
\end{equation}
This is the acceleration equation for the dynamical evolution of the
FRW universe in $(n+1)$-dimensional spacetime. Multiplying
$\dot{a}a$ on both sides of Eq. (\ref{addot}), and using the
continuity equation (\ref{Cont}), after integrating we find
\begin{equation}\label{Fried1}
H^2+\frac{k}{a^2} =\frac{2G_{n+1} }{n-2}\Omega_n
\rho=\frac{4G_{n+1}\pi^{n/2} }{n(n-2)(\frac{n}{2}-1)!} \rho,
\end{equation}
where $k$ is an integration constant. When $n=3$, we have
$\Omega_3=4\pi/3$ and one recovers the standard Friedmann equation
\begin{equation}\label{Friedstand}
H^2+\frac{k}{a^2} =\frac{8\pi G}{3} \rho.
\end{equation}
Is it worth noting that in the literature the $(n+1)$-dimensional
Friedmann equation, in Einstein gravity, usually is written as
\begin{equation}\label{Friedstand2}
H^2+\frac{k}{a^2} =\frac{2\kappa_n }{n(n-1)} \rho,
\end{equation}
with $\kappa_n=8 \pi G_{n+1}$ (sometimes it is also written
$\kappa_n=8 \pi G$). Let us note that the coupling in the r.h.s of
this equation differs from that we derived in Eq. (\ref{Fried1}) for
$n\geq 4$. A question then arises, which one is the correct Einstein
gravitational constant? We believe that the coupling constant we
derived from the entropic force approach is the correct one. To show
this, let us note that the root of the factor $8 \pi$ in Eq.
(\ref{Friedstand}) and also (\ref{Friedstand2}) is the relation
\begin{equation}\label{R00}
R_{00}=\nabla ^2 \phi,
\end{equation}
where $\phi$ is the Newtonian gravitational potential and $R_{00}$
is the (00) component of the Ricci tensor. Now, the coefficient in
the Poisson equation, i.e. $4 \pi$ has been obtained using the Gauss
law for 3-dimensional space. Thus we should first derive the correct
coefficient for $n$-dimensional space. Applying Guass's law for an
$n$-dimensional volume, one finds the Poisson equation for arbitrary
fixed dimension \cite{mansori}
\begin{equation}\label{delphi}
\nabla ^2 \phi=\frac{2G_{n+1}\pi^{n/2} }{(\frac{n}{2}-1)!} \rho,
\end{equation}
On the other hand for $n\geq3$, one finds \cite{mansori}
\begin{equation}\label{Rn}
R_{00}=\left(\frac{n-2}{n-1}\right)\kappa_n \rho,
\end{equation}
Comparing Eqs.  (\ref{R00}), (\ref{delphi}) and (\ref{Rn}) gives us
the following modified Einstein gravitational constant for arbitrary
$n\geq3$ dimensions
\begin{equation}\label{kn}
\kappa_n=\frac{2( n-1)\pi^{n/2} G_{n+1}}{(n-2)(\frac{n}{2}-1)!}.
\end{equation}
Substituting relation (\ref{kn}) in (\ref{Friedstand2}), immediately
shows that the correct form of the Friedmann equation in $n\geq3$
dimension is the expression we derived in Eq. (\ref{Fried1}). This
is a remarkable result and shows that the approach presented here is
powerful enough to derive the correct form of the gravitational
field equations.
\section{ Gauss-Bonnet Entropic Gravity}
Next we study the entropic force idea in Gauss-Bonnet gravity. This
theory contains a special combination of curvature-squared term,
added to the Einstein-Hilbert action. The key point which should be
noticed here is that in Gauss-Bonnet gravity the entropy of the
holographic screen does not obey the area law.  The lagrangian of
the Gauss-Bonnet correction term is given by
\begin{equation}
{\mathcal L}_{GB}=R^2-4R^{ab}R_{ab}+R^{abcd}R_{abcd}.
\end{equation}
The low energy effective action of heterotic string theory naturally
produces the Gauss-Bonnet correction term. The Gauss-Bonnet term
does not have any dynamical effect in four dimensions since it is
just a topological term in four dimensions. Static black hole
solutions of Gauss-Bonnet gravity have been found and their
thermodynamics have been investigated in ample details
\cite{Bou,caigb}. The entropy of the static spherically symmetric
black hole in Gauss-Bonnet theory has the following expression
\cite{caigb}
\begin{equation}\label{Sbh}
S=\frac{k_Bc^3 \Sigma_n}{4\hbar
G_{n+1}}\left[1+\frac{n-1}{n-3}\frac{2
\tilde{\alpha}}{r_{+}^2}\right],
\end{equation}
where $\Sigma_n$ is the horizon area and $r_+$ is the horizon
radius. In the above expression $\tilde{\alpha}=(n-2)(n-3)\alpha$,
where $\alpha$ is the Gauss-Bonnet coefficient which is positive
\cite{Bou}, namely $\alpha>0$. We assume the entropy expression
(\ref{Sbh}) also holds for the apparent horizon of the FRW universe
in Gauss-Bonnet gravity \cite{Cai2}. The only change we need to
apply is the replacement of the horizon radius $r_+$ with the
apparent horizon radius $R$, namely
\begin{equation}\label{S2}
S=\frac{k_B c^3 \Sigma_n}{4\hbar
G_{n+1}}\left[1+\frac{n-1}{n-3}\frac{2
\tilde{\alpha}}{R^2}\right].
\end{equation}
For $n=3$ we have $\tilde{\alpha}=0$, thus the Gauss-Bonnet
correction term contributes only for $n\geq 4$ as we mentioned. In
this case, the change in the entropy  becomes
\begin{equation}\label{deS2}
\triangle S=\frac{k_B c^3 \triangle \Sigma_n }{4\hbar
G_{n+1}}+\frac{n-1}{n-3}\frac{k_B c^3 \tilde{\alpha}}{2\hbar
G_{n+1}} \triangle \left(\frac{\Sigma_n}{R^2}\right) .
\end{equation}
Using the relation $\Sigma_n=n\Omega_n R^{n-1}$ we have $\triangle
\Sigma_n R= (n-1) \Sigma_n \triangle R$. Combining this expression
with Eq. (\ref{deS2}) after using relation $ \triangle \Sigma_n =Q$,
we obtain
\begin{equation}\label{deS3}
\triangle S=\frac{Q k_B c^3 }{4G_{n+1}\hbar}\left[1+\frac{2
\tilde{\alpha}}{R^2}\right] .
\end{equation}
Inserting Eqs. (\ref{T}), (\ref{Q}) and (\ref{deS3}) in Eq.
(\ref{F}) we find
\begin{equation}\label{FGB1}
F=-G_{n+1}\frac{Mm}{R^{n-1}} \left[1+\frac{2
\tilde{\alpha}}{R^2}\right] .
\end{equation}
This is the Newton's law of gravitation in Gauss-Bonnet gravity
resulting from the entropic force approach. In the absence of
Gauss-Bonnet term ($\tilde{\alpha}=0$) one recovers Eq. (\ref{F4}).
It is worth mentioning that the correction term in Eq. (\ref{FGB1})
can be comparable to the first term only when $R$ is very small,
namely for strong gravity. This implies that the correction make
sense only at the very small distances. When $R$ becomes large, i.e.
for weak gravity, the modified Newton's law reduces to the usual
Newton's law of gravitation.

Finally, we derive the $(n+1)$-dimensional Friedmann equation of FRW
universe in Gauss-Bonnet gravity using the approach we developed in
the previous section. In the presence of Gauss-Bonnet term Eq.
(\ref{F7}) is modified as
\begin{equation}\label{FGB2}
\frac{\ddot{a}}{a}=-\frac{2G_{n+1}\pi^{n/2}}{n(\frac{n}{2}-1)!}\rho
\left[1+\frac{2 \tilde{\alpha}}{R^2}\right].
\end{equation}
Note that $R=a(t)r$ is a function of time. Eq. (\ref{FGB2}) is the
dynamical equation for ($n+1$)-dimensional Newtonian cosmology in
Gauss-Bonnet gravity. The main difference between this equation and
Eq. (\ref{F7}) is that the correction term depends explicitly on the
radius $R$. In order to remove this confusion, we suppose that for
Newtonian cosmology the spacetime is Minkowskian with $k=0$. In this
case we have $R=1/H$, and thus we can rewrite Eq. (\ref{FGB2}) as
\begin{equation}\label{FGB2N}
\frac{\ddot{a}}{a}=-\frac{2G_{n+1}\pi^{n/2}}{n(\frac{n}{2}-1)!}\rho
\left[1+2 \tilde{\alpha} \left(\frac{ \dot{a}}{a}\right)^2\right].
\end{equation}
Combining Eq. (\ref{FGB2}) with (\ref{M2}), after replacing $M$ by
$\mathcal M$, we get
\begin{equation}\label{addotGB}
\frac{\ddot{a}}{a} =-\frac{2G_{n+1}\pi^{n/2}
}{n(n-2)(\frac{n}{2}-1)!}[(n-2)\rho+np] \left[1+\frac{2
\tilde{\alpha}}{R^2}\right].
\end{equation}
Thus we have derived the acceleration equation for the dynamical
evolution of the FRW universe in Gauss-Bonnet theory. Multiplying
$\dot{a}a$ on both sides of Eq. (\ref{addotGB}), and using the
continuity equation (\ref{Cont}), we get
\begin{equation}\label{FriedGB0}
d(\dot{a}^2) =\frac{4G_{n+1}\pi^{n/2} }{n(n-2)(\frac{n}{2}-1)!}
\left[d(\rho a^2)+\frac{2 \tilde{\alpha}}{r^2}\frac{d(\rho a^2)}{
a^2}\right].
\end{equation}
Integrating yields
\begin{equation}\label{FriedGB1}
H^2+\frac{k}{a^2} =\frac{4G_{n+1}\pi^{n/2}
}{n(n-2)(\frac{n}{2}-1)!}\rho \left[1+\frac{2 \tilde{\alpha}}{\rho
R^2}\int{\frac{d(\rho a^2)}{a^2}}\right].
\end{equation}
Now, in order to calculate the correction term we need to find
$\rho=\rho(a)$. Suppose a constant equation of state parameter
$w=p/\rho$, integrating the continuity equation (\ref{Cont})
immediately yields
\begin{equation}\label{rho}
\rho=\rho_0 a^{-n(1+w)},
\end{equation}
where $\rho_0$, an integration constant, is the present value of the
energy density. Inserting relation (\ref{rho}) in Eq.
(\ref{FriedGB1}), after integration, we obtain
\begin{equation}\label{FriedGB2}
H^2+\frac{k}{a^2} =\frac{4G_{n+1}\pi^{n/2}
}{n(n-2)(\frac{n}{2}-1)!}\rho \left[1+\frac{2 \tilde{\alpha}}{
R^2}\left(\frac{n(1+w)-2}{n(1+w)}\right)\right].
\end{equation}
Using Eq. (\ref{radius}) we can  further rewrite the above equation
as
\begin{eqnarray}\label{FriedGB3}
&&\left(H^2+\frac{k}{a^2}\right)
\left[1+2\tilde{\alpha}\left(H^2+\frac{k}{a^2}\right)
\frac{n(1+w)-2}{n(1+w)}\right]^{-1}=\frac{4G_{n+1}\pi^{n/2}
}{n(n-2)(\frac{n}{2}-1)!}\rho
\end{eqnarray}
Next, we expand the above equation up to the linear order of
$\tilde{\alpha}$. We find
\begin{eqnarray}\label{FriedGB4}
&&\left(H^2+\frac{k}{a^2}\right)+
{\alpha}^{\prime}\left(H^2+\frac{k}{a^2}\right)^{2}=\frac{4G_{n+1}\pi^{n/2}
}{n(n-2)(\frac{n}{2}-1)!}\rho,
\end{eqnarray}
where we have defined
\begin{eqnarray}\label{alpha}
\alpha ^{\prime}\equiv\frac{2\tilde{\alpha}[2-n(1+w)]}{n(1+w)},
\end{eqnarray}
and we have neglected $O(\tilde{\alpha}^2)$ terms and higher
powers of $\tilde{\alpha}$. This is due to the fact that at the
present time $R\gg 1$ and hence $H^2+k/a^2\ll 1 $. Indeed for the
present time where the apparent horizon radius becomes large, the
correction term is relatively small and the usual Friedman
equation is recovered. Thus, the correction make sense only at the
early stage of the universe where $a\rightarrow 0$. When
$a\rightarrow 0$, even the higher powers of $\tilde{\alpha}$
should be considered. With expansion of the universe, the modified
Friedmann equation reduces to the usual Friedman equation. From
Eq. (\ref{FriedGB4}) we see that the correction term explicitly
depends on the matter content through the equation of state
parameter, $w$, where we have assumed to be a constant.

Eq. (\ref{FriedGB4}) is the $(n+1)$-dimensional Friedmann equation
in Gauss-Bonnet Gravity. The Friedmann equation obtained here from
entropic force approach is in good agreement with that obtained from
the gravitational field equation in Gauss-Bonnet gravity
\cite{Caimy}. This fact further supports the viability of Verlinde
formalism.
\section{ LOVELOCK Entropic Gravity}
Finally we generalize our discussion to a more general case, the
so-called Lovelock gravity, which is a generalization of the
Gauss-Bonnet gravity. The most general lagrangian which keeps the
field equations of motion for the metric of second order, as the
pure Einstein-Hilbert action, is Lovelock lagrangian \cite{Lov}.
This lagrangian is constructed from the dimensionally extended Euler
densities and can be written as
\begin{equation}\label{LL}
L=\sum_{p=0}^{m}c_{p}L_{p},
\end{equation}
where $c_{p}$ and $L_{p}$ are arbitrary constant and Euler density,
respectively. $L_{0}$ set to be one, so $c_{0}$ plays the role of
the cosmological constant, $L_{1}$ and $L_{2}$ are, respectively,
the usual curvature scalar and Gauss-Bonnet term. In an
$(n+1)$-dimensional spacetime $m=[n/2]$. The entropy of the
spherically symmetric black hole solutions in Lovelock theory can be
expressed as \cite{Cai6}
\begin{equation}\label{SL}
S=\frac{k_B c^3\Sigma_{n}}{4\hbar G_{n+1}}\sum_{i=1}^m
\frac{i(n-1)}{n-2i+1}{\hat{c}_{i}}{{r}_{+}}^{2-2i}.
\end{equation}
where $\Sigma_n=n\Omega_{n}r^{n-1}_{+}$ is the horizon area. In the
above expression the coefficients ${\hat{c}_{i}}$ are given by
\begin{equation}\label{constant}
{\hat{c}_{0}}=\frac{{c_{0}}}{n(n-1)}, \ \  {\hat{c}_{1}}=1, \ \
{\hat{c}_{i}}=c_{i}\prod_{j=3}^{2m}(n+1-j) \  \  i>1.
\end{equation}
Note that in expression (\ref{SL}) for entropy, the cosmological
constant term ${\hat{c}_{0}}$ doesn't appear. This is a reasonable
result, and due to the fact that the black hole entropy depends only
on its horizon geometry. We further assume the entropy expression
(\ref{SL}) are valid for a FRW universe bounded by the apparent
horizon in the Lovelock gravity provided we replace the horizon
radius $r_{+}$ with the apparent horizon radius $R$, namely
\begin{equation}\label{mSL}
S=\frac{k_B c^3\Sigma_{n}}{4\hbar G_{n+1}}\sum_{i=1}^m
\frac{i(n-1)}{n-2i+1}{\hat{c}_{i}}R^{2-2i}.
\end{equation}
It is easy to show that, the first term in the above expression
leads to the well-known area law. The second term yields the
apparent horizon entropy in Gauss-Bonnet gravity. The change in the
general entropy expression of Lovelock gravity is obtained as
\begin{equation}\label{deSL}
\triangle S=\frac{k_B c^3 Q }{4\hbar G_{n+1}}\sum_{i=1}^m
i{\hat{c}_{i}}R^{2-2i}.
\end{equation}
where we have used Eq. (\ref{N}). Inserting Eq. (\ref{T}), (\ref{Q})
and (\ref{deSL}) in Eq. (\ref{F}) one finds
\begin{equation}\label{FL1}
F=-G_{n+1}\frac{Mm}{R^{n-1}}\sum_{i=1}^m i{\hat{c}_{i}}R^{2-2i}.
\end{equation}
Thus we have derived the Newton's law of gravitation in Lovelock
gravity resulting from the entropic force. It is obvious that the
first term of the above expression yields the famous Newton's law of
gravity, and the others terms will be important only for strong
gravity or small distances. In this manner, the dynamical equation
for $(n+1)$-dimensional Newtonian cosmology takes the following form
\begin{equation}\label{FL3}
\frac{\ddot{a}}{a}=-\frac{2G_{n+1}\pi^{n/2}}{n(\frac{n}{2}-1)!}\rho
\sum_{i=1}^m i{\hat{c}_{i}}\left(\frac{\dot{a}}{a}\right)^{2i-2}.
\end{equation}
The acceleration equation for the dynamical evolution of the FRW
universe in $(n+1)$-dimensional Lovelock gravity is obtained
following the method developed in the previous section. The result is
\begin{equation}\label{addotL}
\frac{\ddot{a}}{a} =-\frac{2G_{n+1}\pi^{n/2}
}{n(n-2)(\frac{n}{2}-1)!}[(n-2)\rho+np]\sum_{i=1}^m
i{\hat{c}_{i}}R^{2-2i}.
\end{equation}
 Multiplying
$\dot{a}a$ on both sides of Eq. (\ref{addotL}), and using the
continuity equation (\ref{Cont}), after integrating, we get
\begin{equation}\label{FriedL1}
H^2+\frac{k}{a^2} =\frac{4G_{n+1}\pi^{n/2}
}{n(n-2)(\frac{n}{2}-1)!}\rho\left[1+\sum_{i=2}^m\frac{i{\hat{c}_{i}}}{\rho
a^2r^{2i-2}}\int{\frac{d(\rho a^2)}{a^{2i-2}}}\right].
\end{equation}
Using Eq. (\ref{rho}), we can perform the integration. We obtain
\begin{equation}\label{FriedL2}
H^2+\frac{k}{a^2} =\frac{4G_{n+1}\pi^{n/2}
}{n(n-2)(\frac{n}{2}-1)!}\rho\left[1+\sum_{i=2}^m\frac{[2-n(1+w)]i{\hat{c}_{i}}}{
 [2(2-i)-n(1+w)]} \times \frac{1}{R^{2(i-1)}}\right].
\end{equation}
Eq. (\ref{FriedL2}) can be rewritten in the following form
\begin{equation}\label{FriedL3}
\left(H^2+\frac{k}{a^2}\right)\left[1+\sum_{i=2}^m\frac{[2-n(1+w)]i{\hat{c}_{i}}}{
 [2(2-i)-n(1+w)]}\times \frac{1}{R^{2(i-1)}}\right]^{-1}
=\frac{4G_{n+1}\pi^{n/2} }{n(n-2)(\frac{n}{2}-1)!}\rho.
\end{equation}
At the present time where $R\gg1$, we can expand the l.h.s of the
above equation. Using Eq. (\ref{radius}), we reach
\begin{equation}\label{FriedL4}
\left(H^2+\frac{k}{a^2}\right)\left[1-\sum_{i=2}^m\frac{(2-n(1+w))i{\hat{c}_{i}}}{
[2(2-i)-n(1+w)]}\left(H^2+\frac{k}{a^2}\right)^{i-1}\right]
=\frac{4G_{n+1}\pi^{n/2} }{n(n-2)(\frac{n}{2}-1)!}\rho.
\end{equation}
If we define
\begin{equation}\label{betai}
\beta_{i}\equiv\frac{i[2-n(1+w)]{\hat{c}_{i}}}{n(1+w)-2(2-i)},
\end{equation}
 then we can write Eq. (\ref{FriedL4}) in the following form
\begin{eqnarray}\label{FriedL5}
\left(H^2+\frac{k}{a^2}\right)+
\sum_{i=2}^m\beta_{i}\left(H^2+\frac{k}{a^2}\right)^{i}=\frac{4G_{n+1}\pi^{n/2}
}{n(n-2)(\frac{n}{2}-1)!}\rho.
\end{eqnarray}
In this way we derive the $(n+1)$-dimensional Friedmann equations
in Lovelock gravity from the entropic force approach which is
consistent with the result obtained from different methods
\cite{CaiKim,Der}. When $\beta_{i}=0$  ($i\geq2$), one recovers
the standard Friedmann equation in Einstein gravity. The first
term in summation of the above equation is the Gauss-Bonnet
leading correction term derived in the previous section provided
we define $\hat{c}_{2}=\tilde{\alpha}$. In this case $\beta_2$ is
exactly the $\alpha ^{\prime}$ of the previous section. Again,
from Eqs. (\ref{betai}) and (\ref{FriedL5}) we see that the
correction terms explicitly depend on the equation of state
parameter, $w$.

\section{Conclusions and discussions}
According to Verlinde's argument, the total number of bits on the
holographic screen is proportional to the area, $A$, and can be
specified as $N =\frac{A c^3}{G \hbar}$. Indeed, the derivation of
Newton's law of gravity as well as Friedmann equations, in Verlide
formalism, depend on the entropy-area relationship $S=\frac{A
c^3}{4G\hbar}$, where $A=4\pi R^{2}$ represents the area of the
horizon \cite{Ver}. However, it is well known that the area formula
of black hole entropy no longer holds in higher derivative
gravities. So it would be interesting to see whether one can derive
Newton's law of gravity as well as the corresponding Friedmann
equations in these gravities in the framework of entropic force
perspective developed by Verlinde \cite{Ver}.

In this paper, starting from first principles and assuming the
entropy associated with the holographic screen/apparent horizon
given by the expression previously known via black hole
thermodynamics, we were able to derive the Newton's law of gravity
as well as the cosmological equations (Friedmann equations)
governing the evolution of the universe in any gravity theory
including Einstein, Gauss-Bonnet and more general Lovelock gravity
in arbitrary dimensions. We derived the Newton's law of
gravitation from entropic force directly. Then, we derived the
Friedmann equation by equating the mass in Newtonian gravity with
active gravitational mass. Therefore, the Friedmann equation
derived here from entropic force too, but indirectly. In our
derivation the assumption that the entropy of the apparent horizon
of FRW universe in Gauss-Bonnet and Lovelock gravity have the same
form as the spherically symmetric black hole entropy in these
gravities, but replacing the black hole horizon radius by the
apparent horizon radius, plays a crucial role. Interestingly
enough, we found that the higher dimensional gravitational
coupling constant of Friedmann/Einstein equation can be derived
naturally from this approach which coincides with the result
obtained by comparing the weak field limit of Einstein equation
with Poisson equation in higher dimension. Our study shows that
the approach here is powerful enough to derive the gravitational
field equations in any gravity theory. The results obtained here
in the framework of Gauss-Bonnet gravity and more general Lovelock
gravity further support the viability of Verlinde's formalism.

\acknowledgments{This work has been supported by Research Institute
for Astronomy and Astrophysics of Maragha, Iran.}


\begin{thebibliography}{99}
\bibitem{HB} J. D. Bekenstein,  Phys. Rev. D 7,
2333 (1973);\\ S. W. Hawking, Commun Math. Phys. 43, 199 (1975);\\
 S. W. Hawking, Nature 248, 30 (1974).

\bibitem{B} J. M. Bardeen, B. Carter and S. W. Hawking, Commun. Math. Phys. 31,
161 (1973).

\bibitem{D} P. C. W. Davies, J. Phys. A: Math. Gen. 8, 609 (1975);\\
W. G. Unruh, Phys. Rev. D 14, 870 (1976);\\
L. Susskind, J. Math. Phys. 36, 6377 (1995).


\bibitem{Jac} T. Jacobson, Phys. Rev. Lett. {\bf75}, 1260 (1995).

\bibitem{Pad} T. Padmanabhan, Class. Quantum. Grav. 19 (2002)
5387.

\bibitem{Elin} C. Eling, R. Guedens, and T. Jacobson,
Phys. Rev. Lett. {\bf96}, 121301 (2006).

  \bibitem{Cai2} M.~Akbar and R.~G.~Cai, Phys. Rev. D {\bf 75}, 084003 (2007).
  \bibitem{Cai3} R.~G.~Cai and L.~M.~Cao, Phys.Rev. D {\bf 75}, 064008
  (2007).

\bibitem{CaiKim} R. G. Cai and S. P. Kim, JHEP {\bf0502}, 050
(2005).

\bibitem{Wang} B. Wang, E.
Abdalla and R. K. Su, Phys.Lett. B {\bf503},  394 (2001);\\ B. Wang,
E. Abdalla and R. K. Su, Mod. Phys. Lett. A {\bf17},  23 (2002).

\bibitem{Cai33} R. G. Cai, L. M. Cao and Y. P. Hu,  JHEP 0808
(2008) 090.

\bibitem{Shey0} S. Nojiri and S. D. Odintsov, Gen. Relativ. Gravit. 38,
1285 (2006); \\ A. Sheykhi, Class. Quantum Grav. 27 (2010) 025007;
\\ A. Sheykhi, Eur. Phys. J. C  69 (2010)  265.

  \bibitem{Shey1} A. Sheykhi, B. Wang and R. G. Cai, Nucl. Phys. B {\bf
779} (2007)1;\\ R.~G.~Cai and L.~M.~Cao,
  Nucl. Phys. B {\bf785} (2007) 135
  \bibitem{Shey2} A. Sheykhi, B. Wang and R. G. Cai, Phys. Rev. D {\bf
76} (2007) 023515;\\ A. Sheykhi, B. Wang, Phys. Lett. B 678 (2009)
434;\\ A. Sheykhi, JCAP 05 (2009) 019

\bibitem{Pad0} T. Padmanabhan, Rept. Prog. Phys. 73 (2010)
046901.

\bibitem{Ver}E. Verlinde, JHEP 1104, 029 (2011).

\bibitem{Padm} T. Padmanabhan, Mod. Phys. Lett. A 25 (2010) 1129.

\bibitem{Cai4} R.G. Cai, L. M. Cao and N. Ohta, Phys. Rev. D 81, (2010) 061501(R);\\
 R. G. Cai, L. M. Cao and N. Ohta, Phys. Rev. D 81 (2010)
084012.
\bibitem{Smolin} L. Smolin, arXiv:1001.3668.

\bibitem{Li} M. Li and Y. Wang, Phys. Lett. B {687}, 243 (2010).


\bibitem{Tian} Y. Tian and X. Wu, Phys. Rev. D \textbf{81}, 104013 (2010);

Y. S. Myung, arXiv:1002.0871.

\bibitem{Vancea} I. V. Vancea and M. A. Santos, arXiv:1002.2454.


\bibitem{Modesto} L. Modesto and A. Randono, arXiv:1003.1998.

\bibitem{Sheykhi1} A. Sheykhi, Phys. Rev. D \textbf{81}, 104011 (2010).
\bibitem{BLi} B. Liu, Y. C.  Dai, X. R. Hu and J. B. Deng, Mod. Phys. Lett. A {\bf26}, 489 (2011).


\bibitem{Sheykhi2} S. H. Hendi and A. Sheykhi, Phys.  Rev.  D  {\bf83}
(2011) 084012 ;\\ A. Sheykhi and S. H. Hendi, Phys.  Rev.  D {\bf84} (2011) 044023;\\
S. H. Hendi and A. Sheykhi, Int. J Theor. Phys. 51  (2012) 1125;\\
A. Sheykhi, Z. Teimoori, Gen Relativ Gravit 44 (2012) 1129;\\ A.
Sheykhi, Int. J Theor. Phys. 51 (2012) 185.

\bibitem{Gu} W. Gu, M. Li and R. X. Miao,  arXiv:1011.3419;\\ R. X. Miao, J. Meng
and M. Li, arXiv:1102.1166.


\bibitem{other} Y. X. Liu, Y. Q. Wang and S.W. Wei, Class. Quantum
Grav. \textbf{27}, 185002 (2010);\\ V.~V.~Kiselev and
S. A. Timofeev, Mod. Phys. Lett. A {\bf 25}, 2223 (2010);\\
R. A. Konoplya,  Eur.\ Phys.\ J.\  C {\bf 69}, 555 (2010);\\
R. Banerjee and B. R. Majhi. Phys. Rev. D \textbf{81}, 124006 (2010);\\
P. Nicolini, Phys.\ Rev.\  D {\bf 82}, 044030 (2010);\\
C. Gao, Phys. Rev. D \textbf{81}, 087306 (2010);\\  Y. S. Myung
and Y. W. Kim, Phys. Rev. D \textbf{81}, 105012 (2010);\\ H. Wei,
Phys. Lett. B \textbf{692}, 167 (2010);\\  Y. Ling and J.P. Wu,
JCAP \textbf{1008}, 017 (2010);\\ D. A. Easson, P. H. Frampton and
G. F. Smoot, Phys. Lett.B 696 (2011)273;\\
D. A. Easson, P. H. Frampton and G. F. Smoot, arXiv:1003.1528;\\
S. W. Wei, Y. X. Liu and Y. Q. Wang, Commun. Theor. Phys.56 (2011)
455.

\bibitem{mann} R. B. Mann and  J. R. Mureika, Phys. Lett. B 703 (2011)
167.


\bibitem{Lov} D. Lovelock, J. Math. Phys. (N.Y.) 12, 498 (1971).
\bibitem{Der} N. Deruelle and L. Farina-Busto, Phys. Rev. D 41, 3696 (1990).
\bibitem{Zwi} B. Zwiebach, Phys. Lett. B 156, 315 (1985);\\ B. Zumino, Phys.
Rep. 137, 109 (1986);\\ D. J. Gross and J. H. Sloan, Nucl. Phys.
B291, 41 (1987).



\bibitem{Mad} J.M. Maldacena, Adv.
Theor.Math. Phys. 2, 231 (1998);\\ Int. J. Theor. Phys. 38, 1113
(1999).


\bibitem{Sus} L. Susskind, J. Math. Phys. 36, 6377 (1995);\\ R. Bousso, Rev.
Mod. Phys. 74, 825 (2002).



\bibitem{Pad1} T. Padmanabhan, Phys. Rev. D {\bf81}, 124040
(2010).

\bibitem{Pad3} T. Padmanabhan, Class. Quantum Grav., 21, 4485 (2004).
\bibitem{mansori} Reza Mansouri and Ali Nayeri, Gravitation  Cosmology, Vol. 4, No. 2 (1998) 142.


\bibitem{Bou} D. G. Boulware and S. Deser, Phys. Rev. Lett. 55, 2656
(1985);\\
J. T. Wheeler, Nucl. Phys. B 268, 737 (1986);\\ Nucl. Phys. B 273,
732 (1986);\\ R. C. Myers and J. Z. Simon, Phys. Rev. D 38, 2434
(1988).
\bibitem{caigb} R. G. Cai, Phys. Rev. D65, 084014 (2002);\\ R. G. Cai and Q. Guo,
Phys. Rev. D69, 104025 (2004);\\ R. G. Cai and K. S. Soh, Phys. Rev.
D59, 044013 (1999).

\bibitem{Caimy} R. G. Cai and Y. S. Myang, Phys. Lett. B 559 (2003) 60.

\bibitem{Cai6} R. G. Cai, Phys. Lett. B 582, 237(2004).

\end{thebibliography}
\end{document}